\documentclass[%
 aip,
rsi,%
 amsmath,amssymb,
 reprint,%
]{revtex4-1}

\usepackage{graphicx}
\usepackage{dcolumn}
\usepackage{bm}
\usepackage{float}
\usepackage{array}
\usepackage{textcomp}

\begin{document}
\newcolumntype{C}[1]{>{\centering\arraybackslash}p{#1}}
\newcommand{\beginsupplement}{%
        \setcounter{table}{0}
        \renewcommand{\thetable}{S\arabic{table}}%
        \setcounter{figure}{0}
        \renewcommand{\thefigure}{S\arabic{figure}}%
     }
     
\preprint{AIP/123-QED}

\title[Cryogenic microwave loss in epitaxial Al/GaAs/Al trilayers for superconducting circuits]{Cryogenic microwave loss in epitaxial Al/GaAs/Al trilayers for superconducting circuits}

\author{C.R.H. McRae}
\email{coreyrae.mcrae@colorado.edu}
\affiliation{
Department of Physics, University of Colorado, Boulder, CO, USA 80309}
\affiliation{ 
National Institute of Standards and Technology, Boulder, CO, USA 80305
}%
\author{A. McFadden}%
\affiliation{
Department of Electrical and Computer Engineering, University of California, Santa Barbara, CA, USA 93106}
\author{R. Zhao}%
\affiliation{
Department of Physics, University of Colorado, Boulder, CO, USA 80309}
\affiliation{ 
National Institute of Standards and Technology, Boulder, CO, USA 80305
}%
\author{H. Wang}%
\affiliation{
Department of Physics, University of Colorado, Boulder, CO, USA 80309}
\affiliation{ 
National Institute of Standards and Technology, Boulder, CO, USA 80305
}%
\author{J.L. Long}%
\affiliation{
Department of Physics, University of Colorado, Boulder, CO, USA 80309}
\affiliation{ 
National Institute of Standards and Technology, Boulder, CO, USA 80305
}%
\author{T. Zhao}%
\affiliation{
Department of Physics, University of Colorado, Boulder, CO, USA 80309}
\affiliation{ 
National Institute of Standards and Technology, Boulder, CO, USA 80305
}%
\author{S. Park}%
\affiliation{
Department of Physics, University of Colorado, Boulder, CO, USA 80309}
\affiliation{ 
National Institute of Standards and Technology, Boulder, CO, USA 80305
}%
\author{M. Bal}%
\affiliation{
Department of Physics, University of Colorado, Boulder, CO, USA 80309}
\affiliation{ 
National Institute of Standards and Technology, Boulder, CO, USA 80305
}%
\author{C.J. Palmstrøm}%
\affiliation{
Department of Electrical and Computer Engineering, University of California, Santa Barbara, CA, USA 93106}
\affiliation{
Materials Department, University of California, Santa Barbara, CA, USA 93106}
\author{D.P. Pappas}%
\affiliation{ 
National Institute of Standards and Technology, Boulder, CO, USA 80305
}%

\date{\today}

\begin{abstract}
Epitaxially-grown superconductor/dielectric/superconductor trilayers have the potential to form high-performance superconducting quantum devices and may even allow scalable superconducting quantum computing with low-surface-area qubits such as the merged-element transmon. In this work, we measure the power-independent loss and two-level-state (TLS) loss of epitaxial, wafer-bonded, and substrate-removed Al/GaAs/Al trilayers by measuring lumped element superconducting microwave resonators at millikelvin temperatures and down to single photon powers. The power-independent loss of the device is $(4.8 \pm 0.1) \times 10^{-5}$ and resonator-induced intrinsic TLS loss is $(6.4 \pm 0.2) \times 10^{-5}$. Dielectric loss extraction is used to determine a lower bound of the intrinsic TLS loss of the trilayer of $7.2 \times 10^{-5}$. The unusually high power-independent loss is attributed to GaAs's intrinsic piezoelectricity.

%
\end{abstract}

\maketitle

\section{\label{sec:level1}Introduction}

The investigation of the electrical properties of dielectric materials and interfaces in the millikelvin-temperature and single-photon-power regime is a burgeoning field in superconducting microwave circuits and is critical to performance enhancement in superconducting quantum computing.~\cite{McRae2020b} In particular, epitaxially-grown dielectrics are of interest because crystalline materials with low defect density have the potential to exhibit lower two-level-system (TLS) loss,~\cite{Oh2005,Nakamura2011} the dominant form of loss in high performance superconducting quantum circuits.~\cite{Muller2019,McRae2020b} In addition, the ultra-high vacuum environment used in epitaxial growth allows for lower TLS loss attributed to cleaner interfaces between materials.~\cite{Megrant2012,Richardson2016} 

The discovery of a low-loss superconductor/dielectric/superconductor trilayer would allow the implementation of scalable, high-performance quantum computing designs such as the merged-element transmon.~\cite{Zhao2020}

Because the epitaxial growth of GaAs and Al/GaAs heterostructures is well-established,~\cite{Cho1978,Petroff1981,Ludeke1981,Pilkington1999} GaAs is a natural candidate for epitaxial growth for superconducting quantum devices.

In this work, we measure the power-independent loss and TLS loss of epi-Al/GaAs/Al trilayers on $\mathrm{Al_2O_3}$ made using a wafer bonding technique.~\cite{McFadden2020} To determine this loss, we perform cryogenic microwave measurements of lumped element superconducting microwave resonators with parallel plate capacitors formed from these trilayers. We demonstrate that these epitaxial films perform similarly to bulk GaAs~\cite{Scigliuzzo2020} in both high- and low-power regimes, and exhibit loss dominated by power-independent loss which we attribute to the intrinsic piezoelectricity of GaAs.

\section{Device Design and Fabrication}

\begin{figure}
\includegraphics[width=80mm]{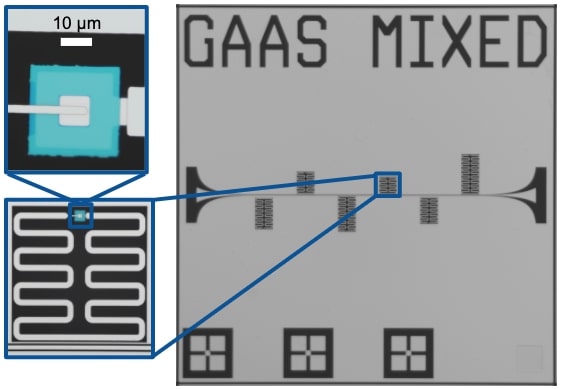}
\caption{\label{fig:res_image}Optical micrographs of a lumped element resonator with a Al/GaAs/Al parallel plate capacitor and liftoff Al inductor. Blue region is GaAs, light grey is Al, and dark grey is sapphire substrate. Blue squares show a zoom-in of the trilayer resonator with an inductor design of N = 7 (seven inductor meanders). Chip shown is sample Trilayer 1, with a zoom-in of device A, as in Table~\ref{tab:resmeas}.}
\end{figure}

\begin{table*}
\caption{\label{tab:resmeas}Parameters extracted from cryogenic microwave measurements of lumped element resonators with Al/GaAs/Al parallel plate capacitors (Trilayer) and interdigitated capacitors (Planar). All measurements were performed in DR1 unless stated otherwise. Values are given with their 95$\%$ confidence intervals where available. $N$: number of inductor meanders. $f_0$: resonance frequency. $1/Q_{i,\mathrm{HP}}$: inverse high power internal quality factor. $F \delta_{\mathrm{TLS}}^{0}$: resonator-induced intrinsic TLS loss. $1/Q_{i,\mathrm{LP}}$: inverse low power internal quality factor. This value is reported if $F \delta_{\mathrm{TLS}}^{0}$ is unavailable. $1/Q_c$: inverse coupling quality factor. Resonator A (B) TLS fit determined critical number of photons $n_c$ = 50,000 (70,000) and exponential constant $\beta$ = 3 (0.37).}
\begin{ruledtabular}
\begin{tabular}{ccccccccc}
Device Label & Sample & $N$ & $f_0$ (GHz) & $1/Q_{i,\mathrm{HP}}$ ($\times 10^{-6}$) & $F \delta_{\mathrm{TLS}}^{0}$ ($\times 10^{-6}$) & $1/Q_{i,\mathrm{LP}}$ ($\times 10^{-6}$) & $1/Q_c$ ($\times 10^{-6}$) \\
\hline
A & Trilayer 1 & 7 & 7.41 & 48 $\pm$ 1 & 64 $\pm$ 2  & - & 15.3 \\
B & Planar 1 & 7 & 7.92  & 0.3 $\pm$ 0.1 & 6.2 $\pm$ 0.2 & - & 2.6 \\
\hline
-& Trilayer 1 DR2 & 17 & 4.79 & 57.8 & $\sim$ 18.3 & - & 23.6 \\
-& Trilayer 1 & 9 & 6.39 & 108 & - & $\sim$ 217 & 17 \\
-& Trilayer 2 & 7 & 7.41 & 92.6 & - & $\sim$ 110 & 163
\end{tabular}
\end{ruledtabular}
\end{table*}

\begin{figure}
\includegraphics[width=85mm]{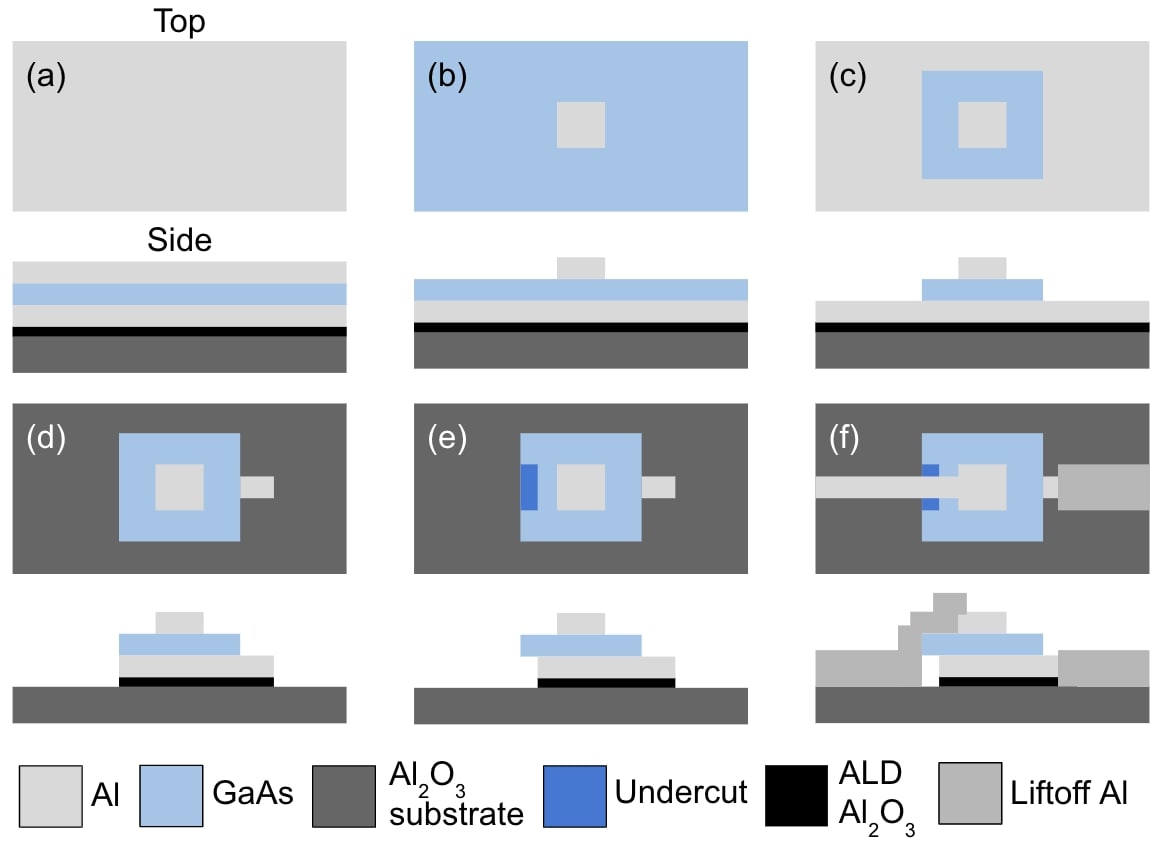}
\caption{\label{fig:gaas_fab}Diagram of fabrication process for epi-Al/GaAs/Al trilayer lumped element resonators, with top and side views of capacitor. (a) shows trilayer prior to processing. Fabrication steps shown are (b) capacitor top plate definition, (c) GaAs layer etch, (d) bottom Al layer etch, (e) undercut, and (f) Al liftoff.
}
\end{figure}

The material under test is a 40 nm epi-Al/40 nm epi-GaAs/40 nm epi-Al trilayer with a 20 nm atomic-layer-deposited (ALD) $\mathrm{Al_2O_3}$ bonding layer, bonded to an $\mathrm{Al_2O_3}$(0001) substrate. The interfaces are abrupt and epitaxial, and the GaAs is single crystalline as determined by transmission electron microscopy (TEM).~\cite{McFadden2020} More details on the growth, wafer bonding, substrate removal, and regrowth processes, as well as materials imaging and characterization, can be found in Ref.~\citenum{McFadden2020}.
Trilayer lumped element resonators (shown in Fig.~\ref{fig:res_image}) are patterned using a six-step lithography and etch process (Fig.~\ref{fig:gaas_fab}). First, the top capacitor plate is defined using Megaposit MF26A developer~\footnote{Certain commercial equipment, instruments, and materials are identified in this paper to foster understanding. Such identification does not imply recommendation or endorsement by the National Institute of Standards and Technology, nor does it imply that the materials or equipment identified are necessarily the best available for the purpose.} to etch the top layer of Al (Fig.~\ref{fig:gaas_fab}~(b)). Then, the GaAs is etched away with Transene GA300 wet etchant heated to 33 $\mathrm{^\circ C}$ (Fig.~\ref{fig:gaas_fab}~(c)). A second MF26A etch is used to remove the next layer of Al as well as the $\mathrm{AlO_x}$ bonding layer (Fig.~\ref{fig:gaas_fab}~(d)). An undercut of the bottom capacitor plate is performed with Transene D Al Etchant (Fig.~\ref{fig:gaas_fab}~(e)). Transene A Al Etchant is used to fully remove residual Al in large blank areas. Finally, a liftoff process of e-beam deposited Al is used to form the feedline and inductors (Fig.~\ref{fig:gaas_fab}~(f)). Auto-spun Megaposit SPR660 photoresist exposed using a maskless aligner is used for lithography in all but the liftoff step, where a trilayer of MicroChem PMMA A2, MicroChem LOR5A, and SPR660 are used. An oxygen plasma ash is used to prepare the surface prior to Al deposition.

The trilayer resonator design is similar to that in Ref.~\citenum{McRae2020}. The parallel plate capacitor is designed to have a 10 $\mathrm{\mu m}$ $\times$ 10 $\mathrm{\mu m}$ top plate and is connected to each end of the inductor by liftoff. The inductor is 15 $\mathrm{\mu m}$ in width with a gap between inductive meanders of 30 $\mathrm{\mu m}$ and inductor length varies by varying the number of meanders $N$ between 7 and 17, corresponding to resonance frequencies $f_0$ between 4.7 and 7.5 GHz. Coupling quality factors vary between roughly 10,000 and 100,000 in order to facilitate critical coupling.

About 17$\%$ of the total Al top electrode area is liftoff Al, not epi-Al. This could obscure the epi-Al/GaAs/Al loss if trilayer loss is much lower than liftoff interface loss. In addition, the capacitor undercut has an estimated participation of 6$\%$ based on the capacitance of that region. In future experiments, the undercut connection will be replaced by an airbridge in order to increase measurement sensitivity to the trilayer loss.

In order to take into account the effect of the inductor circuit element, trilayer resonator measurements are compared to those of planar resonators, for which the same inductor design is used but the trilayer is replaced by an interdigitated capacitor. Planar resonators are fabricated using liftoff e-beam Al on sapphire to imitate the inductor fabrication in the trilayer devices.

\begin{figure*}
\includegraphics[width=140mm]{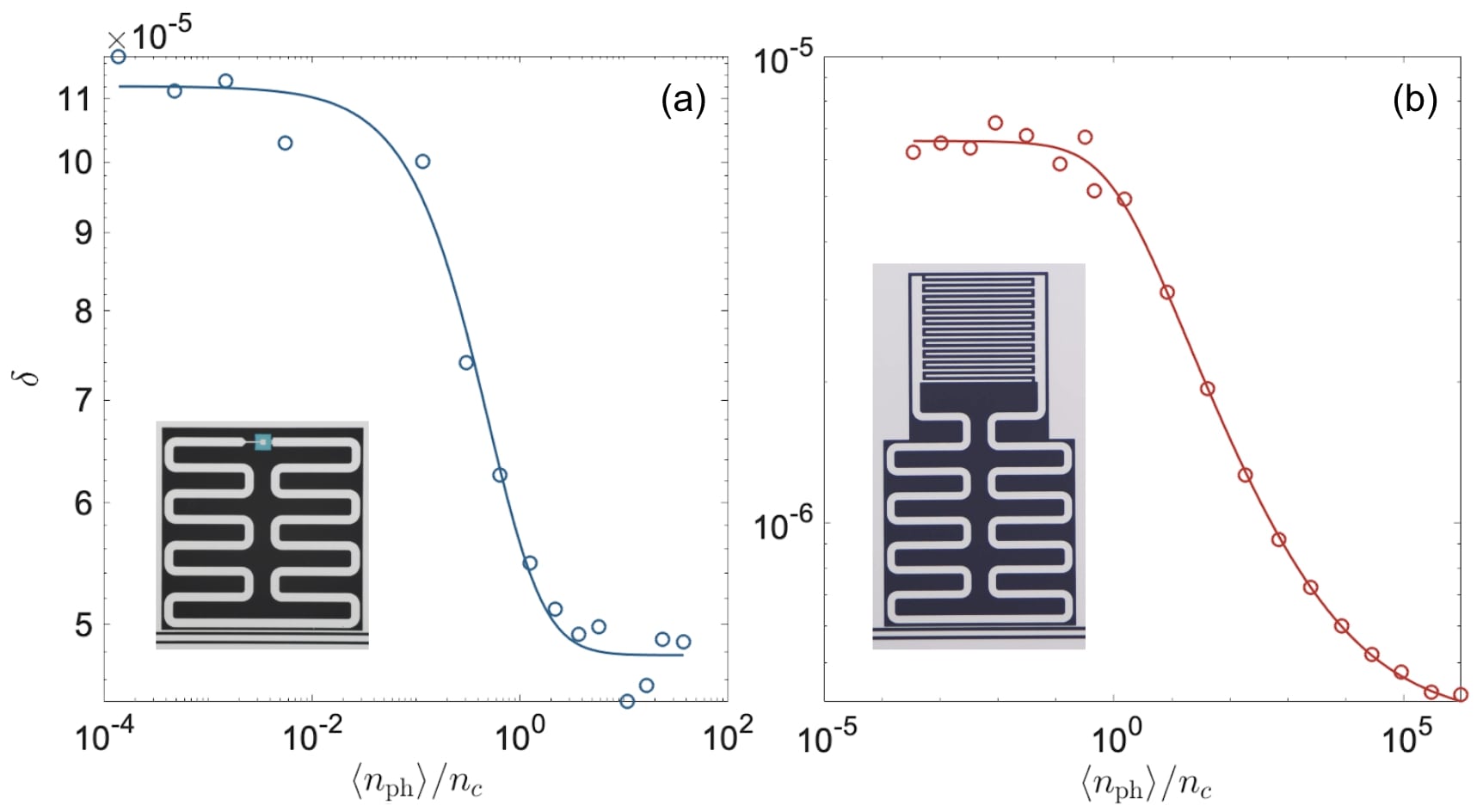}
\caption{\label{fig:powersweep}Loss $\delta$ as a function of normalized number of photons in the resonator $\langle n_{\textrm{ph}} \rangle / n_c$ at T $\sim$ 12 mK for resonators (a) A and (b) B, defined in Table~\ref{tab:resmeas}. Data is denoted as circles, and fitting to the total loss model in Eq.~\ref{eqn:totallossmodel} is denoted as a solid line. Insets show representations of the measured devices.
}
\end{figure*}

\section{Microwave Measurements and Loss Extraction}

Samples are clamped into sample boxes made from gold-plated oxygen-free high thermal conductivity copper, with wirebonds used for electrical connection. Measurements are performed in two different cryogen-free dilution refrigerators (DRs): DR1, at a temperature of 12 mK, and DR2, at a temperature below 10 mK. Applied power is varied between -5 and -90 dBm with roughly 70 dB of line attenuation. 

Internal quality factors $Q_i$, coupling quality factors $Q_c$, and resonance frequencies $f_0$ are determined by way of a fitting routine that implements the diameter correction method,~\cite{Khalil2012} with a fixed-$f_0$ Monte Carlo fit, ten points on either side of the resonance frequency used for normalization, and one 3-dB bandwidth of data around the resonance used for the fitting itself. Resonator data, as well as measurement and fitting codes, can be found online.~\footnote{\lowercase{h}ttps://github.com/Boulder-Cryogenic-Quantum-Testbed/}

Four trilayer resonators and one planar resonator were successfully measured, as shown in Table~\ref{tab:resmeas}, and two were able to be fitted at sufficiently high and low powers as to allow TLS model fitting.~\cite{Pappas2011} These two devices are labeled device A (trilayer resonator) and device B (planar resonator). Measurements of the other trilayer resonators in Table~\ref{tab:resmeas} support the values seen in the measurements of device A.

Fig.~\ref{fig:powersweep} shows power sweeps of loss $\delta = 1/Q_i$ for resonators A and B as well as fits to the
total loss model
\begin{equation}
\label{eqn:totallossmodel}
\delta = \delta_{TLS}(T, \langle n_{\textrm{ph}} \rangle) + \delta_{\mathrm{other}}(T),
\end{equation}
where $\delta_{TLS}$ is TLS loss as defined in the following equation and varies with temperature $T$ as well as time-averaged number of photons in the resonator $\langle n_{\textrm{ph}} \rangle$, and $\delta_{\mathrm{other}}$ is a sum of power-independent losses such as quasiparticle loss, vortex loss, and piezoelectric loss.~\cite{Scigliuzzo2020} The TLS model is given by~\cite{Burnett2018,Pappas2011}
\begin{equation}
\label{eqn:TLSlossmodel}
\delta_{TLS} = F \delta_{TLS}^0 \frac{\tanh(\frac{\hbar \omega}{ 2 k_B T})}{(1 + \frac{\langle n_{\textrm{ph}} \rangle}{n_c})^\beta},
\end{equation}
where $\delta^0_{TLS}$ is the intrinsic TLS loss, $n_c$ is the resonator's critical photon number, $\beta$ is an exponential constant describing the homogeneity of the TLS population,
and $F$ is the filling factor of the TLS material. The resonator-induced intrinsic TLS loss, $F \delta^0_{TLS}$, is the effective loss due to TLS in the low-power, low-temperature limit. TLS model fitting results for Fig.~\ref{fig:powersweep} are reported in Table~\ref{tab:resmeas}. $\langle n_{\textrm{ph}} \rangle$ is estimated using the resonator $Q_c$, $Q_i$, and $f_0$, and total power, as in Ref.~\citenum{Burnett2018}.

At high power, where TLS are saturated and power-independent losses dominate, $\delta = 1/Q_{i,HP} \sim \delta_{other}$ where $1/Q_{i,HP}$ is the inverse high power internal quality factor. In this regime, the loss of trilayer resonator A is $(4.8 \pm 0.1) \times 10^{-5}$. This value is only a factor of two different than the loss at low powers (Fig.~\ref{fig:powersweep}~(a)), demonstrating that power-independent loss sources dominate the total loss of this device. Indeed, this high-power loss is more than an order of magnitude higher than expected, as shown by comparison to a power sweep of resonator B (Fig.~\ref{fig:powersweep}~(b)), which does not contain an Al/GaAs/Al trilayer. This unusually high loss in the high-power regime is demonstrated repeatably with four trilayer devices on two chips across multiple cooldowns, as shown in Table~\ref{tab:resmeas}.

The high-power loss in this work is within a factor of 3 of the piezoelectric loss measured in bulk GaAs.~\cite{Scigliuzzo2020} We believe that this unusually high high-power loss can also be attributed to piezoelectricity. Quasiparticle and vortex loss are unlikely to vary so significantly between samples measured using the same experimental set-up, such as resonators A and B. Another possible materials loss that would present at high-power is loss due to interdiffusion between the Al and GaAs epitaxial layers. This can be ruled out by interface TEM images in Ref.~\citenum{McFadden2020} that show these interfaces as abrupt and epitaxial.

The resonator-induced intrinsic TLS loss in resonator A is $(6.4 \pm 0.2) \times 10^{-5}$. We can determine a lower bound for the intrinsic TLS loss of the Al/GaAs/Al trilayer, independent of the effect of the resonator wiring, by implementing a modified version of dielectric loss extraction.~\cite{McRae2020} Details on the simulations used can be found in the supplementary material for Ref.~\citenum{McRae2020}. The resonator-induced intrinsic TLS loss in device A, $\delta_{A} = F_A \delta_{\mathrm{TLS},A}^0$, where $F_A$ is the filling factor of the TLS material in device A, is a weighted sum of intrinsic TLS loss in the planar inductor $\delta_L$ and the Al/GaAs/Al trilayer capacitor $\delta_{\mathrm{Al/GaAs/Al}}$, as:
\begin{equation}
    \delta_{A} = \frac{C_L}{C_{\mathrm{tot}}}\delta_{L} + \frac{C_C}{C_{\mathrm{tot}}}\delta_{\mathrm{Al/GaAs/Al}},
\end{equation}
where $C_L$ ($C_C$) is the capacitance of the inductor (capacitor) circuit component, and total capacitance is $C_{\mathrm{tot}} = C_L + C_C$. If we assume all TLS loss in planar resonator B is from the inductor, which is identical in design to the inductor in device A, then $\delta_{B} = \delta_{L}$. We can then determine a lower bound on the loss of the epitaxial Al/GaAs/Al trilayer by
\begin{equation}
\label{eq:trilayerloss}
    \delta_{\mathrm{Al/GaAs/Al}} = \frac{C_{\mathrm{tot}}}{C_C} (\delta_{A} - \frac{C_L}{C_{\mathrm{tot}}}\delta_{B})
\end{equation}
with trilayer capacitance $C_C = 285$ fF and inductor capacitance $C_L = 37.5$ fF, as determined by analysis and simulation, and loss values shown in Table~\ref{tab:resmeas}. For a parallel plate capacitor, $F = 1$, so we can say $\delta_{\mathrm{Al/GaAs/Al}} = \delta^0_{\mathrm{TLS},{\mathrm{Al/GaAs/Al}}}$.

From Eq.~\ref{eq:trilayerloss}, we determine that $\delta_{\mathrm{Al/GaAs/Al}} = 7.2 \times  10^{-5}$, slightly higher than the resonator-induced TLS loss for device A, $\delta_{A} = (6.4 \pm 0.2) \times 10^{-5}$, verifying that the TLS loss of device A is dominated by the Al/GaAs/Al trilayer. This value includes the loss of the undercut region, the liftoff Al on the capacitor, and capacitor edge effects, which could increase the effective loss. Even so, these values agree within a factor of two with bulk GaAs TLS loss measurements.~\cite{Scigliuzzo2020}

For a target qubit lifetime of 50 to 100 $\mathrm{\mu s}$, losses must fall within the mid-$10^{-7}$ range. Thus, the measured loss in this materials set in both the high- and low-power regimes is too high for superconducting qubit applications. However, power-independent and TLS losses may be lower in other similar epitaxial trilayers such as Al/Si/Al and Al/Ge/Al, which will be explored in future work.

\section{Conclusion and Next Steps}

Due to the presence of significant power-independent loss in thin epitaxial Al/GaAs/Al trilayers, GaAs can be ruled out as a promising dielectric material for superconducting quantum computing applications unless mitigation methods are implemented. In the future, similar growth, fabrication and measurement techniques could be applied to other promising, non-piezoelectric materials sets such as epi-Al/Si/Al trilayers, and could yield substantial performance enhancement. 

In future materials measurements of this type, the replacement of the undercut connection with an airbridge could reduce loss from this region and increase measurement sensitivity. In addition, the variation of capacitor size and thickness could allow the extraction of losses for individual regions within the capacitor such as the superconductor/dielectric interfaces and the superconductor and dielectric surfaces. 

\section*{Data Availability}
The data that support the findings of this study are openly available in Boulder-Cryogenic-Quantum-Testbed/data at http://doi.org/10.5281/zenodo.4025406.

\begin{acknowledgments}
We wish to acknowledge the partial support of the Army Research Office, Google, the NIST Quantum Initiative, and the National Science Foundation Grant Number 1839136.
\end{acknowledgments}

\bibliography{GaAsLoss}

\end{document}